\newtheorem{definition}{Definition}
\newtheorem{theorem}{Theorem}
\newcommand{\bt}{\begin{theorem}}
\newcommand{\et}{\end{theorem}}
\newcommand{\bd}{\begin{definition}}
\newcommand{\ed}{\end{definition}}
\newcommand{\be}{\begin{equation}}
\newcommand{\ee}{\end{equation}}
\newcommand{\bear}{\begin{eqnarray}}
\newcommand{\eear}{\end{eqnarray}}
\newcommand{\baar}{\begin{array}}
\newcommand{\eaar}{\end{array}}
\newcommand{\nn}{\nonumber}
\newcommand{\pr}{\partial}
\newcommand{\G}{\Gamma}
\newcommand{\thesection}{\Roman{section}}
\newcommand{\thesubsection}{\Alpha{subsection}}
\documentclass[11pt]{article}
\usepackage{latexsym}
\usepackage{amsfonts}
\usepackage[dvips]{graphicx}

\begin{document}
\renewcommand{\thesection}{\Roman{section}.}
\renewcommand{\thesubsection}{\Alph{subsection}.}
\thispagestyle{empty}
\setlength{\topmargin}{-20mm}
\setlength{\textheight}{250mm}
{\begin{center}
\large{\bf The Riemann-Lanczos Problem as an Exterior Differential System 
           with Examples in 4 and 5 Dimensions}
\end{center}}
\vspace{0.5cm}\noindent
P Dolan$^{\: a)}$\\
Mathematics Department, Imperial College, 180 Queen's Gate,\\
London SW7 2BZ
\newline
\vspace{0.5cm}
\newline
A Gerber$^{\: b)}$\\
Centre for Techno-Mathematics and Scientific Computing Laboratory,\\
University of Westminster, Watford Road, Harrow HA1 3TP
\vspace{1.0cm}
\indent
\begin{center} 
\large{Abstract} 
\end{center}

The key problem of the theory of exterior differential systems (EDS) is to 
decide whether or not a system is in involution. The special case of EDSs
generated by one-forms (Pfaffian systems) can be adequately illustrated by a 
2-dimensional example.

In 4 dimensions two such problems arise in a natural way, namely, the 
Riemann-Lanczos and the Weyl-Lanczos problems. It is known from the work of 
Bampi and Caviglia that the Weyl-Lanczos problem is always in involution in 
both 4 and 5 dimensions  but that the Riemann-Lanczos problem fails to be in
involution even for 4 dimensions. However, {\it singular} solutions of it can
be found.

We give examples of singular solutions for the G\"{o}del, 
Kasner and Debever-Hubaut spacetimes. It is even possible that the singular
solution can inherit the spacetime symmetries as in the Debever-Hubaut case.
We comment on the Riemann-Lanczos problem in 5 dimensions which is neither
in involution nor does it admit a 5-dimensional involution of Vessiot vector 
fields in the generic case.
\vspace{0.3cm}
\footnoterule
\vspace{0.3cm}\noindent
$^{a)}$ Electronic mail: pdolan@inctech.com\\
$^{b)}$ Electronic mail: a\_gerber01@hotmail.com
\newpage
\indent
\setlength{\arraycolsep}{0.2cm}
\setlength{\textheight}{250mm}
\setlength{\topmargin}{-20mm}
\section{Introduction}
The problem of generating the spacetime Weyl conformal curvature 
tensor $C_{abcd}$ from a tensor potential is called the Weyl-Lanczos problem 
and the analogous problem for the Riemann curvature tensor the Riemann-Lanczos 
problem.

The Lanczos tensor potential admits the following index symmetries
\be
   L_{[ab]c}=L_{abc} \: , \label{1antis}
\ee
where $a,b,c,s = 0,1,2,3$ and
\be
   L_{[abc]}=0 \label{1cyclic} \: .
\ee
Apart from these, we may impose two gauge conditions: the differential gauge
condition
\be
{L_{ab}}^{s}_{\: ;s}=0 \: ,\label{1diffg}
\ee
where $'';''$ indicates covariant differentiation, and the algebraic gauge or 
trace free condition
\be
{{L_{a}}^{s}}_{s}=0 \label{1trace} \: .
\ee

Lanczos discovered the Weyl-Lanczos equations \cite{Lanc}, where he 
introduced a Lagrangian based on the double dual of the 
Riemann tensor $R_{abcd}$. The Lanczos tensor $L_{abc}$ arose as a 
Lagrange multiplier for this Lagrangian. Lanczos found an 
expression for the Weyl tensor in terms of certain Lagrange multipliers 
$L_{abc}$, namely,
\bear
C_{abcd} & = &
L_{abc;d}-L_{abd;c}+L_{cda;b}-L_{cdb;a}
+g_{bc}L_{(ad)}+g_{ad}L_{(bc)}-g_{bd}L_{(ac)} \nn\\
& & -g_{ac}L_{(bd)}+\frac{2}{3}{L^{ms}}_{m;s}(g_{ac}g_{bd}-g_{ad}g_{bc}) 
\: ,\label{1Weyll}
\eear
where $L_{ad}={{L_{a}}^{s}}_{d;s}-{{L_{a}}^{s}}_{s;d}$. These are {\it the 
Weyl-Lanczos equations}. They can also be expressed in the form
\be
C_{abcd} =L_{[ab][c;d]} + L_{[cd][a;b]}- ^{*}L^{*}_{[ab][c;d]} 
- ^{*}L^{*}_{[cd][a;b]}\label{1WeylDD}
\ee
as in \cite{Rob1}. Further, if we impose (\ref{1diffg}) and 
(\ref{1trace}) above, we can simplify (\ref{1Weyll}) considerably to
\be
C_{abcd} = L_{abc;d}-L_{abd;c}+L_{cda;b}-L_{cdb;a}-g_{bc}L^{s}_{ad;s}
-g_{ad}L^{s}_{bc;s}+g_{bd}L^{s}_{ac;s}+g_{ac}L^{s}_{bd;s} \: .
\label{1WeylS}
\ee
Many solutions to (\ref{1WeylS}) are known and solutions for vacuum spacetimes
can be found in \cite{DoMu}.
 
We can also attempt to express the Riemann curvature tensor in terms of 
a comparable tensor potential $\hat{L}_{abc}$ which leads to the 
Riemann-Lanczos problem for spacetimes and which has been discussed at 
length by Bampi and Caviglia \cite{Bam1,Bam2}. Udeschini Brinis 
\cite{Ude1} had wanted to describe the spacetime Riemann tensor in terms 
of a Lanczos tensor $\hat{L}_{abc}$ and proposed the Riemann-Lanczos 
relations 
\be
R_{abcd} = \hat{L}_{abc;d}-\hat{L}_{abd;c}+\hat{L}_{cda;b}
-\hat{L}_{cdb;a} \: . \label{1Rieml}
\ee 
But the difficulties with the relations (\ref{1Rieml}) were pointed out in
two papers by Bampi and Caviglia \cite{Bam1,Bam2} where they proved existence 
theorems for solutions of (\ref{1Weyll}) and (\ref{1Rieml}). Whereas the 
Weyl-Lanczos problem is always in involution, the Riemann-Lanczos problem is 
{\it not} and only singular solutions of it can occur if the problem is 
{\bf not} modified. Bampi and Caviglia showed that:\\
i)  The Weyl-Lanczos problem (\ref{1Weyll}) or (\ref{1WeylS}) has 
non-singular solutions for $n=4,5$, where $n$ is the dimension of the
space-time manifold $M$;\\ 
ii) For $n=4$ the Riemann-Lanczos problem (\ref{1Rieml}) has no non-singular 
solutions but it does have ``singular'' solutions which means that the Cartan 
characters do not adopt their maximal values;\\
iii) The differential gauge condition (\ref{1diffg}) has no effect on the 
existence or non-existence of solutions of either (\ref{1Weyll}) or 
(\ref{1Rieml}).

Bampi and Caviglia \cite{Bam2} also suggested a prolongation of the
Riemann-Lanczos equations to a second-order system which makes them 
a system in involution.

In the Riemann-Lanczos problem, we meet equation (\ref{1cyclic}) and possibly 
(\ref{1diffg}) but {\bf not} (\ref{1trace}), which leaves us with 20
independent components for the $\hat{L}_{abc}$ in 4 dimensions.
{\it We always assume that the cyclic conditions} (\ref{1cyclic}) {\it hold} 
but {\bf not} {\it the trace-free conditions} (\ref{1trace}). 
If equations (\ref{1trace}) were to hold, we would have $$ 
R = 4 {\hat{L}^{nk}}_{\:\:\:\: k;n} = -4 {\hat{L}^{nk}}_{\:\:\:\: n;k} = 0 
\: ,$$
which would lead to inconsistencies. Because we are only going to talk about 
the Riemann-Lanczos problem in this paper, we shallchange notation from
$\hat{L}_{abc}$ to $L_{abc}$ from now on. We shallalso write the 
Riemann-Lanczos equations in solved form as
\be
f^{(R)}_{abcd} := R_{abcd} - L_{abc;d} + L_{abd;c}-L_{cda;b}
+L_{cdb;a} \: . \label{Riemso}
\ee
The paper by Massa and Pagani \cite{Mass} concerning a modification of the 
Riemann-Lanczos problem used a totally different approach to
the above work of Bampi and Caviglia. Accordingly, we do not 
consider \cite{Mass} here. As reference \cite{Idiot} is based on 
\cite{Mass} it is also not applicable here.
\section{Exterior Differential Systems and Cartan Characters}

It is necessary to introduce some notions from the theory of exterior 
differential systems (EDS) in order to understand the structure of the
Riemann-Lanczos problem. We denote a formal $N$-dimensional manifold by 
$\mathcal{M}$ of which the spacetime manifold $M$ is an $n$-dimensional 
submanifold. Then, for a collection of differential forms we define an 
exterior differential system (EDS).
\bd
An {\bf exterior differential system} EDS $\Sigma$ on $\mathcal{M}$ consists of
\[ \baar{ll}
1-forms \quad \alpha^{(1)}_{i_{1}} \quad & (1 \leq i_{1} 
\leq k_{1}) \: ,\\
2-forms \quad \alpha^{(2)}_{i_{2}} \quad & (1 \leq i_{2} 
\leq k_{2}) \: ,\\
\quad \vdots & \\
p-forms \quad \alpha^{(p)}_{i_{p}} \quad & (1 \leq i_{p} 
\leq k_{p}) \: ,\\
\eaar \]
and sometimes 0-forms on $\mathcal{M}$.
\ed
Any EDS $\Sigma$ is a subset of the set of all differential forms on 
$\mathcal{M}$. There is a further set of differential forms on $\mathcal{M}$ 
formed by a subset of $\Sigma$ called its {\bf associated ideal} 
$\mathcal{I}(\Sigma)$ \cite{5man,Yang}. 
The associated ideal $\mathcal{I}(\Sigma)$ is a {\bf differential ideal 
$\mathcal{I}$}, that is, $d \mathcal{I} \subset \mathcal{I}$. This means 
that such an ideal is closed under exterior differentiation. If the exterior
differential system (EDS) $\Sigma$ itself is closed under exterior 
differentiation, we say that $\Sigma$ is a {\it closed exterior differential
system}. Next, we wish to look at Pfaffian systems and Vessiot vector fields
as they are of great importance to our applications later on.
\subsection{Pfaffian Systems and Vessiot Vector Fields}

A {\bf Pfaffian system} $\mathcal{P}$ is a special EDS containing only 
one-forms $\theta^{\alpha}$ and 0-forms. We denote the {\bf rank} of 
$\mathcal{P}$ by $s$ which is given by the number $s$ of independent 
one-forms in $\mathcal{P}$. 
Consider now the collection of all vector fields on $\mathcal{M}$, 
$\mathcal{X}(\mathcal{M})$. There is a subset of $\mathcal{X}(\mathcal{M})$ 
given by all those vector fields which annihilate the Pfaffians in
$\mathcal{P}$ $$\mathcal{D}:= \{X \in \mathcal{X}(\mathcal{M})| 
\theta^{\alpha}(X)=0, \quad \alpha = 1, \cdots, s\}$$
so that the number of independent vector fields in $\mathcal{D}$ is $N-s$.
We call $\mathcal{D}$ the {\bf dual system} to $\mathcal{P}$. 
For any Pfaffian system $\mathcal{P}$, we can also look at its {\bf derived
system} $\mathcal{P}'$ for which we can write \cite{Stor}
$$ \mathcal{P}':=\{\theta^{\alpha} \in \mathcal{P}| 
d \theta^{\alpha}(X,Y)=0, \quad \forall X,Y \in \mathcal{D}, \quad 
\alpha = 1, \cdots , s \} \: ,$$
where $\mathcal{P}' \subset \mathcal{P}$ always holds.

Pfaffian systems are classified according to the ease with which they can 
be integrated. The most familiar Pfaffian Systems are also the simplest, 
namely, the
{\bf complete Pfaffian systems} to which the celebrated {\bf Frobenius theorem}
applies. We say that a Pfaffian system $\mathcal{P}$ is {\it complete}
if $\mathcal{P} = \mathcal{P}'$ holds. A necessary and sufficient condition
for a complete Pfaffian system is now given by the Frobenius theorem.
\bt{Frobenius Theorem}\\
Let the $s$ independent 1-forms $\theta^{1},...,\theta^{s}$ generate a
closed ideal $\mathcal{I}(\mathcal{P})$, then, we can find coordinates
in a neighbourhood of any point $x^{1},...,x^{N}$ such that all forms 
are generated by the $s$ coordinate differentials $d x^{N-s+1}, \cdots ,
d x^{N}$.
\et
A good account of Pfaffian systems can be found in \cite{Gar3,Awan} or in a 
more theoretical approach in \cite{Free}. {\it In general, Pfaffian systems 
are not complete} but each Pfaffian system can be enlarged by adding further 
1-forms until the enlarged system is complete or the enlarged system becomes
inconsistent. The minimal enlarged Pfaffian system which is complete is known 
as the {\bf associated system} $\mathcal{A}(\mathcal{P})$ of $\mathcal{P}$ 
and its dimension is called the {\bf class} $c$ of $\mathcal{P}$. Sometimes, 
$\mathcal{A}(\mathcal{P})$ is also called the {\bf Cartan system} 
$C(\mathcal{P})$ \cite{Kamr}.

A Pfaffian system $\mathcal{P}$ can dually be characterised using 
{\bf Vessiot vector field systems} as discussed in \cite{Stor}.
This approach was developed in Vessiot \cite{Vess} and can be found in 
\cite{Fack}. First, we take a vector field system $\mathcal{V}$, which we 
here choose to be the dual system $\mathcal{D}$ of a Pfaffian system 
$\mathcal{P}$ so that $\mathcal{V}=\mathcal{D}$ is given by $N-s$ vector 
fields $Y^{i}$ and we define in turn its {\bf first and second derived
systems} as
\bear
\mathcal{D}' & := & \mathcal{D} + [\mathcal{D},\mathcal{D}]\nn\\
\mathcal{D}'' & := & \mathcal{D'} + [\mathcal{D'},\mathcal{D'}] \nn\\
 & \vdots & 
\eear
and so on \cite{Stor}. 
For the derived system, the inclusion $\mathcal{D} \subset 
\mathcal{D}'$ is always valid and $(\mathcal{P}')^{\perp}=\mathcal{D}'$ for a 
Pfaffian system, where $\mathcal{D}=\mathcal{P}^{\perp}$.
We can also formulate the notion of completeness in terms of vector field 
systems and versions of the Frobenius theorem in this language are numerous.
We therefore define the space $\bar{\mathcal{D}} \subset \mathcal{D}$ as
$$\bar{\mathcal{D}}:=\{ Y \in \mathcal{D} | d \theta^{\alpha}(X,Y)=0 \: 
\forall X \in \mathcal{D}, \: \alpha = 1, \cdots , s \} \: .$$ 
Here, we state the dual version of the Frobenius theorem as
\bt{Complete Pfaffian Systems}\\
Given any two vector fields $X$, $Y$ $\in \mathcal{D}$, then 
$\mathcal{P}$ is a {\bf complete Pfaffian system} means that the
the commutator vector field $[X,Y]$ is also a vector field in $\mathcal{D}$.
\et
A Vessiot vector field system $\mathcal{D}$ is complete means then that 
$[\mathcal{D},\mathcal{D}] \subseteq \mathcal{D}$ for all $Y \in \mathcal{D}$ 
which is equivalent to $\bar{\mathcal{D}} =\mathcal{D}$.

When a vector field system $\mathcal{D}$ or Pfaffian system $\mathcal{P}$ fails
to be complete, a slightly weaker condition may hold on some subsystem of 
$\mathcal{D}$. Such a possibility is that a vector field system is in 
{\bf involution}, which is defined as follows:
\bd{Involutory Subsystem of a Vector Field System}\\
An involutory subsystem $\mathcal{T}$ of a vector field system $\mathcal{D}$ 
is a subsystem $\mathcal{T}$ of $\mathcal{D}$ such that 
$[\mathcal{T},\mathcal{T}] \subseteq \mathcal{D}$, which means that 
$\mathcal{T}$ is closed relative to $\mathcal{D}$ but not relative to 
$\mathcal{T}$ itself.
\ed
This means that if we are given two vector fields $X, Y \in \mathcal{T}$, it 
is then valid for all $\theta^{\alpha} \in \mathcal{P}$
\bear
d \theta^{\alpha} (X,Y) & = & \theta^{\alpha}(X)Y-\theta^{\alpha}(Y)X
-\frac{1}{2}\theta^{\alpha}([X,Y]) \nn\\
& = & -\frac{1}{2}\theta^{\alpha}([X,Y]) \nn\\
& \neq & 0 \: .
\eear
This is because we have $\theta^{\alpha}(X)=0$ and $\theta^{\alpha}(Y)=0$
holding identically {\bf but not} $\frac{1}{2}\theta^{\alpha}([X,Y])=0$ since 
$[X,Y]$ need not lie inside $\mathcal{T}$. However, if we allow for the vector
fields in $\mathcal{D} - \mathcal{T}$, then $\theta^{\alpha}([X,Y])$ can 
vanish i.e.
\bear
d \theta^{\alpha}(X,Y) & = & 0 \: ( \mbox{mod} \: \mathcal{J}(\mathcal{P})) 
\: .
\eear  
Note that when we take $\mathcal{T}=\mathcal{D}$, we get a complete vector
field system but when $\mathcal{T}$ is strictly contained in $\mathcal{D}$ 
it can fail to be complete \cite{Stor}.

Further, the dual space of $C(\mathcal{P})$ which is usually denoted by 
$C(\mathcal{D})=C(\mathcal{P})^{\perp}$ is the 
space of all Cauchy characteristic vector fields, where such a vector field is 
defined as
\bd{Cauchy Characteristic Vector Field}\\
Y is a Cauchy characteristic vector field of a vector field ideal $\mathcal{I}
(\mathcal{P})$ means that 
$$Y \rfloor \theta^{\alpha} =0$$ 
$$Y \rfloor d \theta^{\alpha} 
= \lambda_{\gamma}\theta^{\gamma} = 0 \: (\mbox{mod} \: \mathcal{P})$$
for all one-forms $\theta^{\alpha}$ in $\mathcal{P}$, where 
$\lambda_{\gamma}$ are scalar multipliers for each $\theta^{\alpha}$. 
\ed
Both $C(\mathcal{D})$ and $C(\mathcal{P})$ are complete 
Pfaffian systems and it is $C(\mathcal{D})=\bar{\mathcal{D}}$. For their 
dimensions it has to hold $\mbox{dim}(C(\mathcal{D}))+\mbox{dim}
(C(\mathcal{P}))= N$. Cauchy characteristic vector fields can also be 
characterised as \cite{Pri1}
\bt
Given a Pfaffian system $\mathcal{P}$ and its associated ideal
$\mathcal{I}(\mathcal{P})$ and let $\mathcal{D}$ be the dual space of 
$\mathcal{P}$. $Y \in \mathcal{D}$ is Cauchy means that
$ [X,Y] \in \: \mathcal{D} \quad \forall X \in \mathcal{D}$. 
\et

For EDS and Pfaffian systems in particular, it can be of interest to 
determine its {\bf symmetries} and especially their infinitesimal generators. 
In the same way as Killing vector fields are symmetries of the metric tensor 
and hence of Einstein's equations, any EDS or system of PDEs can adopt 
symmetries. When an EDS adopts symmetries it means that there exist vector 
fields $Y$ whose Lie derivatives applied to each form in the EDS annihilates 
them or those new forms are contained in the EDS itself. These infinitesimal 
symmetry generators $Y$ are called {\bf isovectors} and defined as
\bd{Isovectors}\\
Y is an isovector of an EDS $\Sigma$ means that $\pounds_{Y}\Sigma \subset 
\Sigma$ and dually $Y$ is an isovector of a vector field system  $\mathcal{D}$ 
means that $\pounds_{Y} \mathcal{D} \subset \mathcal{D}$.
\ed
Cauchy characteristics form a special sub-algebra of all isovectors of a 
system\cite{Pri1}. Isovectors $Y$ do not necessarily have to be such 
that $Y \in \mathcal{D}$. Often there exist isovectors $Y$ such that 
$Y \notin \mathcal{D}$. But for those $Y$ with $Y \in D$, it is known that
\cite{Pri1}
\bt{Cauchy Characteristics}\\
Given a vector field ideal $\mathcal{I}(\mathcal{P})$. Then, 
$Y \in \mathcal{D}$ is an isovector means that $Y$ is a Cauchy 
characteristic vector field.
\et
Note that if $\mathcal{P}$ is complete, then all its vector fields $Y \in
\mathcal{D}(=\bar{\mathcal{D}})$ are isovectors (= Cauchy characteristics).
The formula of H. Cartan 
\be
\pounds_{Y}\alpha = Y \rfloor d \alpha + d (Y \rfloor \alpha)
\ee
is used when we determine the isovectors of p-forms $\alpha$. See 
\cite{Blu1,Barc,Edel,Ste1,Olv1} for discussions of symmetries and isovectors. 
An algebraic computing package assisting with symmetry determination is the 
REDUCE package DIMSYM by Sherring \cite{Pri2}.

We illustrate the above theory with the example of a 
coordinate transformation to Euclidean coordinates in two dimensions.
For the line element we therefore have
$$ d s^{2} = y^{2}d x^{2} + d y^{2} =  d u^{2} + d U^{2} \: , $$
where $u, \: U$ are the Euclidean coordinates. For this to hold we must have
\bear
y^{2} & = & p^{2} + P^{2} \: , \nn\\
0 & = & p q + P Q \: ,\nn\\
1 & = & q^{2} + Q^{2} \: , \label{2cord2} 
\eear
where we introduced the Monge notation
$$p:= \frac{\pr u}{\pr x}, \: 
  q:= \frac{\pr u}{\pr y}, \: 
  r:= \frac{\pr^{2} u}{\pr x^{2}}, \: 
  s:= \frac{\pr^{2} u}{\pr x \pr y}, \:
  t:= \frac{\pr^{2} u}{\pr y^{2}} \: ,$$

$$P:= \frac{\pr U}{\pr x}, \: 
  Q:= \frac{\pr U}{\pr y}, \: 
  R:= \frac{\pr^{2} U}{\pr x^{2}}, \: 
  S:= \frac{\pr^{2} U}{\pr x \pr y}, \:
  T:= \frac{\pr^{2} U}{\pr y^{2}} \: .
\label{2coord} $$
These local coordinates form a jet bundle $\mathcal{J}^{2}(\mathbb{R}^{2},
\mathbb{R}^{2})$ with $N=14$ formal dimensions, where we have $n=2$ 
independent and $m=2$
dependent variables. After differentiating equations (\ref{2cord2}) with 
respect to $x$ and $y$ and rearranging them, we obtain a Pfaffian system 
$\mathcal{P}$ with $s=12$ independent 1-forms which can locally be expressed as
\bear
\theta^{1} & = & d u - p d x - q d y \nn\\
\theta^{2} & = & d U - P d x - Q d y \nn\\
\theta^{3} & = & d p - r d x - s d y \nn\\
\theta^{4} & = & d P - R d x - S d y \nn\\
\theta^{5} & = & d q - s d x - t d y \nn\\
\theta^{6} & = & d Q - S d x - T d y \nn\\
\theta^{7} & = & p d r + r d p + P d R + R d P \nn\\
\theta^{8} & = & q d r + r d q + Q d R + R d Q + d y \nn\\
\theta^{9} & = & p d s + s d p + P d S + S d P - d y \nn\\
\theta^{10} & = & q d s + s d q + Q d S + S d Q \nn\\
\theta^{11} & = & p d t + t d p + P d T + T d P \nn\\
\theta^{12} & = & q d t + t d q + Q d T + T d Q \: . \label{2Pfacor}
\eear
As the rank of $\mathcal{D}$ is $N-s=14-12=2$, the system (\ref{2Pfacor}) can 
dually be characterised by two Vessiot vector fields $V^{1}, V^{2}$ 
generating $\mathcal{D}$, where 
\bear
V^{i} & = & V^{i}_{x} \frac{\pr}{\pr x}+V^{i}_{y} \frac{\pr}{\pr y}
+(p V^{i}_{x}+q V^{i}_{y})\frac{\pr}{u}+(P V^{i}_{x}+Q V^{i}_{y})
\frac{\pr}{\pr U}+(r V^{i}_{x} \nn\\
& & +s V^{i}_{y})\frac{\pr}{\pr p} 
+(R V^{i}_{x}+S V^{i}_{y})\frac{\pr}{\pr P}
+(s V^{i}_{x}+t V^{i}_{y})\frac{\pr}{\pr q} \nn\\
& & +(S V^{i}_{x}+T V^{i}_{y})\frac{\pr}{\pr Q} 
+V^{i}_{r}\frac{\pr}{\pr r}+V^{i}_{R}\frac{\pr}{\pr R}
+V^{i}_{s}\frac{\pr}{\pr s}+V^{i}_{S}\frac{\pr}{\pr S} \nn\\
& & +V^{i}_{t}\frac{\pr}{\pr t}+V^{i}_{T}\frac{\pr}{\pr T} \: ,
i = 1,2 \label{2Pfvess} 
\eear
and the coefficients $V^{i}_{r},V^{i}_{R},V^{i}_{s},V^{i}_{S},V^{i}_{t},
V^{i}_{T}$ are determined through the condition that the six 1-forms 
$\theta^{7}$ to $\theta^{12}$ are annihilated by $V^{1},V^{2}$. Therefore, we 
obtain
\bear
V^{i}_{r} & = & \frac{1}{\alpha}[(\frac{P}{Q}(rs+RS)-r^{2}-R^{2})V^{i}_{x}
+(\frac{P}{Q}(rt+RT+1)-rs-RS)V^{i}_{y}] \: ,\nn\\
V^{i}_{R} & = & -\frac{1}{Q}[(rs+RS+\frac{q}{\alpha}(\frac{P}{Q}
(rs+RS-r^{2}-R^{2}))V^{i}_{x}+(rt+RT+1 \nn\\
& & +\frac{q}{\alpha}(\frac{P}{Q}(rt+RT+1)-rs-RS)V^{i}_{y}] \: , \nn\\
V^{i}_{s} & = & \frac{1}{\alpha}[(\frac{P}{Q}(s^{2}+S^{2})-rs-RS)V^{i}_{x}
+(\frac{P}{Q}(st+ST)-s^{2}-S^{2}+1)V^{i}_{y}] \: , \nn\\
V^{i}_{S} & = & -\frac{1}{Q}[(s^{2}+S^{2}+\frac{q}{\alpha}(\frac{P}{Q}
(s^{2}+S^{2})-rs-RS))V^{i}_{x}+(st+ST \nn\\
& & +\frac{q}{\alpha}(\frac{P}{Q}(st+ST)-s^{2}-S^{2}+1)V^{i}_{y}] \: , \nn\\
V^{i}_{t} & = & \frac{1}{\alpha}[(\frac{P}{Q}(st+ST)-rt-RT)V^{i}_{x}
+(\frac{P}{Q}(t^{2}+T^{2})-st-ST)V^{i}_{y}] \: , \nn\\
V^{i}_{T} & = & -\frac{1}{Q}[(st+ST+\frac{q}{\alpha}(\frac{P}{Q}
(st+ST)-rt-RT))V^{i}_{x} \nn\\
& & +(t^{2}+T^{2}+\frac{q}{\alpha}(\frac{P}{Q}
(t^{2}+T^{2})-st-ST))]V^{i}_{y} \: , \label{2Vescom}
\eear
where $\alpha$ is given by $\alpha = pQ-Pq$. Equations (\ref{2Vescom}) tell 
us that all components of $V^{1}, V^{2}$ are determined by $V^{i}_{x},
V^{i}_{y}$ which can be assigned arbitrarily. In this way all 2-forms 
arising from (\ref{2Pfacor}) and applied to $V^{1},V^{2}$ vanish identically. 
This means that the system (\ref{2Pfacor}) is {\it complete} and its derived 
system consists of the two Vessiot vector fields $V^{1}, V^{2}$ spanning 
$\mathcal{D}$ so that $\mathcal{D}= \bar{\mathcal{D}}=\mathcal{D}'$. It also 
means that $C(\mathcal{P})=\mathcal{D}$ so that $c=2$ and therefore, both 
$V^{1}$ and $V^{2}$ are Cauchy charcateristics. At the same time, they form 
an involution of maximal dimension $g=2$. Note that the $V^{1},V^{2}$ are also
isovectors but that other isovectors which are not in $\mathcal{D}$ such as 
$Y^{i}=C^{i}_{u}\frac{\pr}{\pr u}+C^{i}_{U}\frac{\pr}{\pr U}\: n=1,2$ do occur.
\subsection{Integral Elements, Cartan Characters and Genus}

All definitions in this section are based on \cite{Moli} which relies on 
Cartan's original work whereas the work in \cite{Kaeh} varies slightly from 
\cite{Cart} in some definitions. When an EDS $\Sigma$ or a Pfaffian 
system $\mathcal{P}$ is given, we try to find manifolds on which 
all the differential forms of $\Sigma$ or $\mathcal{P}$ are annihilated. 
Firstly, we shallconstruct tangent spaces to such manifolds, where the 
dimension of each of them can be deduced from a sequence of non-negative 
integers called the Cartan characters.

Assume, we are given a vector space $\mathbb{R}^{N}$ of which we want to find
all possible distinct p-dimensional subspaces. The set of such subspaces forms
a $p(N-p)-$dimensional manifold called the {\bf Grassmann manifold}
$G(N,p)$. We can characterise $G(N,p)$ as a quotient space of orthogonal groups
as follows:
$$G(N,p) \simeq O(N,p) / O(N) \: \mbox{{\bf x}} \: O(p) \: .$$
Alternatively, one can write each $p$-dimensional vector space as a 
{\it p-vector} $v= \lambda e^{1} \wedge \cdots \wedge e^{p}$, where $\{ e^{1}, 
\cdots , e^{p} \}$ is its basis. Using this definition we obtain
$$G(N,p) \simeq \{ (v^{1}, \cdots , v^{p} \}/_{\sim} \: ,$$
where $(v^{1}\wedge \cdots \wedge v^{p}) \sim (w^{1} \wedge \cdots \wedge 
w^{p}) \Leftrightarrow (w^{1}, \cdots , \cdots w^{p}) =\lambda (v^{1}, 
\cdots , v^{p})$ meaning that all multiples of a $p$-vector are in the same
equivalence class. The {\bf Pl\"{u}cker coordinate} of a $p$-vector 
$v=\frac{1}{p!} a_{i_{1}}\cdots a_{i_{p}} e^{i_{1}}\wedge \cdots \wedge 
e^{i_{p}}$ is given by the equivalence class $[(a_{i_{1}}\cdots a_{i_{p}})]$ 
of which each equivalence class represents a distinct $p$-plane.

Because we are not only looking at a single $G(N,p)$ but at a whole set for a 
given manifold $\mathcal{M}$, we define the {\bf Grassmann bundle as}
$$G_{p}(\mathcal{M}) := \bigcup_{x \in {\mathcal{M}}} G(N,p)|_{x} \: ,$$
where $G(N,p)|_{x}$ is the Grassmannian manifold at a point 
$x \in \mathcal{M}$. We shalluse these definitions later on to help to 
describe integral manifolds but first we introduce integral elements as
\bd{Integral Element}\\
A p-dimensional subspace $(E^{p})_{x}$ of the tangent space 
$T_{x}(\mathcal{M})$ at a point $x$ on a $N$-dimensional manifold 
$\mathcal{M}$ is called an integral element of an EDS 
$\Sigma$ if $\alpha^{i}_{j}(E^{p})_{x}=0$ at $x$ for any form 
$\alpha^{i}_{j}$ in $\Sigma_{p}$, $ 1 \leq i \leq p, \: 1 \leq j \leq
k_{p},$ which means that all differential forms of $\Sigma$ are annihilated on 
$(E^{p})_{x}$ at $x$ $\in \mathcal{M}$.
\ed
Each $p$-dimensional integral element $(E^{p})_{x}$ is then a particular 
element of $G(N,p)|_{x}$. The set of all $p$-dimensional integral elements
$(E^{p})_{x}$ for all $x \in \mathcal{M}$ is then denoted by 
$\mathcal{V}_{p}(\Sigma)$, where $\mathcal{V}_{p}(\Sigma) \subset 
G_{p}(\mathcal{M})$.

Once we have found such a $p$-dimensional integral element $(E^{p})_{x} 
\subset T_{x}\mathcal{M}$ spanned by $V^{1}, \cdots ,V^{p}$, we
are looking for vectors $V^{p+1}_{x} \in T_{x} \mathcal{M}$ in such a way 
that the space generated by $(E^{p})_{x}$ and $V^{p+1}_{x}$ form a 
$(p+1)$-dimensional integral element. The conditions on such a tangent 
vector $V^{p+1}_{x}$ at the point $x \in \mathcal{M}$ are:
\bear
\alpha^{1}_{i_{1}}(V^{p+1}_{x}) & = & 0, \quad 1 \leq i_{1} \leq k_{1} \: , 
\nn\\
\alpha^{2}_{i_{2}}(V^{j_{1}},V^{p+1}_{x}) & = & 0, \quad 1 \leq i_{2} 
\leq k_{2}\: , 1 \leq j_{1} \leq p \: , \nn\\
 & & \vdots \nn\\
\alpha^{p+1}_{i_{p+1}}(V^{1},V^{2}, \cdots ,V^{p},V^{p+1}_{x}) 
& = & 0, \quad 1 \leq i_{p+1} \leq k_{p+1} \label{2Polar} \: .
\eear
The vectors $V^{p+1}_{x}$ that satisfy the above {\bf polar system 
$H((E^{p})_{x})$} of linear equations generate the {\bf polar space 
$H((E^{p})_{x})^{\perp}$} of $(E^{p})_{x}$. Depending on the ranks of the 
polar systems generated, we can divide integral elements into 3 classes which 
we call {\bf regular}, {\bf ordinary} and {\bf singular}. We look at the 
subsystem $\alpha^{1}_{1}(V^{1}_{x}) = \cdots = \alpha^{1}_{k_{1}}(V^{1}_{x}) 
= 0$ of (\ref{2Polar}) where $s_{0}(x)$ is its rank. From this, we get the 
integer $$s_{0}:= max_{\{ x \in \mathcal{M}\} }(r(H((E^{0})_{x}))=s_{0}(x))$$ 
called the {\bf ${zeroth}$ Cartan character}. We define a {\bf regular 
point} $x \in \mathcal{M}$ to be a point where $s_{0}(x)=s_{0}$. Then, a 
1-dimensional integral element $(E^{1})_{x}$ is an {\bf ordinary} integral 
element if $x$ is a regular point. Let the polar system of $(E^{1})_{x}$ in 
$T_{x}\mathcal{M}$ have rank $r(H((E^{1})_{x}))=s_{1}(x)+s_{0}$. From this, 
we define
$$s_{1}:= max_{\{x \in \mathcal{M}, V^{2} \in T_{x}\mathcal{M}\}}
r(H((E^{1})_{x}))-s_{0},$$
which is called the {\bf first Cartan character}. 
If, for $x \in \mathcal{M}, \: s_{1}(x)=s_{1}$ holds, then the integral 
element $(E^{1})_{x}$ is called {\bf regular}. A 2-dimensional integral 
element $(E^{2})_{x}$ then is called {\bf ordinary} if it contains at least 
one regular 1-dimensional integral element. We define inductively
\bd{Cartan Characters}\\
The $p^{th}$ Cartan character is inductively defined as:\\
$s_{p}=max_{\{ x \in \mathcal{M}, V^{p+1} \in T_{x}\mathcal{M} \}}
 r(H((E^{p})_{x}))-\sum_{i=0}^{p-1}s_{i}$.
\ed
Using the notion of Cartan characters we can now precisely define
\bd{Regular, Ordinary and Singular Integral Elements}\\
A p-dimensional integral element $(E^{p})_{x}$ is called {\bf ordinary} if it
contains at least one (p-1)-dimensional regular integral element. 
A p-dimensional integral element is called {\bf regular} if its polar system
$H((E^{p})_{x})$ has maximal rank $r(H((E^{p})_{x}))=s_{p}(x)=s_{p}$. 
A p-dimensional integral element is {\bf singular} if it is neither regular
nor ordinary.
\ed
Note that a sequence of integral elements $(E^{0})_{x} \subset (E^{1})_{x} 
\subset \cdots \subset (E^{p})_{x}$ at a point $x \in \mathcal{M}$ is called a
{\bf regular chain} of integral elements if all its $(E^{k})_{x}, \quad 
1 \leq k \leq p-1,$ are regular and $(E^{p})_{x}$ is at least ordinary.
The maximal dimension an ordinary integral element can adopt is called the
{\bf genus} $g$ of $\Sigma$ or $\mathcal{P}$. 

Integral elements are the tangent planes to manifolds which are solution 
manifolds for a given EDS $\Sigma$ and Pfaffian systems $\mathcal{P}$ and 
defined as
\bd{Integral Manifolds}\\
An integral manifold of an EDS $\Sigma$ on $\mathcal{M}$ is a
p-dimensional submanifold $\mathcal{N}$ of $\mathcal{M}$ such that each 
k-dimensional vector subspace $(E^{k})_{x} \subset T_{x}\mathcal{N}$ for 
$1 \leq k \leq p$ annihilates all the k-forms in the EDS $\Sigma$.
\ed
Next, we wish to illustrate this theory using example (\ref{2Pfacor}) again.
There, a 1-dimensional integral element $(E^{1})_{x}$ can for instance be 
created by means of setting $V^{1}_{x}:=1, \: V^{1}_{y}:=0$ 
and computing the other components according to (\ref{2Pfvess}) and 
(\ref{2Vescom}). Its polar space is then spanned by $H((E^{1})_{x})= 
\{ V^{2} \}$ for any $V^{2}$ satisfying (\ref{2Pfvess}) and 
(\ref{2Vescom}) other than $V^{1}$ itself. Then, $(E^{2})_{x}$ is given by 
$(E^{2})_{x}=\{ V^{1}, \: V^{2}\}$ which is the maximal dimension an 
integral element can adopt for this example, where we can for instance choose
$V^{2}_{x}:=0$ and $V^{2}_{y}:=1$. The polar matrix of the polar space
$H(E^{0})_{x})$ is given by
\[ \left( \baar{ll|llllllllllll}
d x & d y & d u & d U & d p & d P & d q & d Q & d r & d R & d s & d S & 
d t & d T \nn\\ \hline
-p & -q & 1 & 0 & 0 & 0 & 0 & 0 & 0 & 0 & 0 & 0 & 0 & 0 \nn\\
-P & -Q & 0 & 1 & 0 & 0 & 0 & 0 & 0 & 0 & 0 & 0 & 0 & 0 \nn\\
-r & -s & 0 & 0 & 1 & 0 & 0 & 0 & 0 & 0 & 0 & 0 & 0 & 0 \nn\\
-R & -S & 0 & 0 & 0 & 1 & 0 & 0 & 0 & 0 & 0 & 0 & 0 & 0 \nn\\
-s & -t & 0 & 0 & 0 & 0 & 1 & 0 & 0 & 0 & 0 & 0 & 0 & 0 \nn\\
-S & -T & 0 & 0 & 0 & 0 & 0 & 1 & 0 & 0 & 0 & 0 & 0 & 0 \nn\\
0 & 0 & 0 & 0 & r & R & 0 & 0 & p & P & 0 & 0 & 0 & 0 \nn\\
0 & 1 & 0 & 0 & 0 & 0 & r & R & q & Q & 0 & 0 & 0 & 0 \nn\\
0 & -1 & 0 & 0 & s & S & 0 & 0 & 0 & 0 & p & P & 0 & 0 \nn\\
0 & 0 & 0 & 0 & 0 & 0 & s & S & 0 & 0 & q & Q & 0 & 0 \nn\\
0 & 0 & 0 & 0 & t & T & 0 & 0 & 0 & 0 & 0 & 0 & p & P \nn\\
0 & 0 & 0 & 0 & 0 & 0 & t & T & 0 & 0 & 0 & 0 & q & Q 
\eaar \right) \: . \]
The rank $r(H(E^{0})_{x})$ of the polar matrix is equal to the rank $s$ of
our Pfaffian system (\ref{2Pfacor}) and therefore we obtain $s_{0}=s=12$.  
The two Vessiot vector fields $V^{i}$ which form a 
2-dimensional involution for (\ref{2Pfacor}) also constitute a regular chain 
of integral elements where $(E^{1})_{x}$ is spanned by either of the $V^{i}$ 
and $(E^{2})_{x}$ by both of them.

Next, we wish to determine the integral manifold for (\ref{2Pfacor})
on which (\ref{2cord2}) holds as well.
From the 6 equations obtained from differentiating (\ref{2cord2}) with 
respect to $x$ and $y$, we see that $t=\frac{\pr^{2} u}{\pr y^{2}}=0$ and
$T=\frac{\pr^{2} U}{\pr y^{2}}=0$ so that the solution must be linear in $y$
\bear
u & = & y f(x)+g(x) \nn\\
U & = & y F(x) +G(x) 
\eear
with $f(x),g(x),F(x),G(x)$ arbitrary functions. From this we find by very 
elementary calculations that the general solution is given by
\[ \baar{ll} \label{2Param}
u = y \cos{(x+\epsilon)}+a \: , & 
U = y \sin{(x+\epsilon)}+b \: , \nn\\
 & \nn\\
p = -y \sin{(x+\epsilon)} \: , & 
P = y \cos{(x+\epsilon)} \: , \nn\\
 & \nn\\
q = \cos{(x+\epsilon)} \: , & 
Q = \sin{(x+\epsilon)} \: , \nn\\
 & \nn\\
r = - y \cos{(x+\epsilon)} \: , & 
R = y \cos{(x+\epsilon)} \: , \nn\\
 & \nn\\
s = -\sin{(x+\epsilon)} \: , & 
S = \cos{(x+\epsilon)} \: , \nn\\
 & \nn\\
t = 0 \: , & T = 0 \: , 
\eaar \]
where $\{a,b, \epsilon\}$ are arbitrary constants corresponding to the 
translational and rotational degrees of freedom. This also gives a local
parameterisation of the 2-dimensional integral manifold which corresponds to 
this solution.
\subsection{Independence Condition, Reduced Characters and Involution}

Until now, all variables of a given EDS $\Sigma$ were treated equally. 
But if we wish to look for specific integral manifolds transversal to 
some given submanifold, we must specify this fact for our EDS. This is 
achieved by introducing an {\bf independence condition} $\Omega=\omega^{1} 
\wedge \cdots \wedge \omega^{n} \neq 0$, which is sometimes also called the 
{\it volume element}, where the $\omega^{i}$ are 1-forms 
characterising such a submanifold. If we select $x^{1}, \cdots , x^{n}$ as
local coordinates, where we shalluse brackets to indicate powers of $x^{i}$
such as in ${(x^{i})}^{n}$, then $\Omega$ is given by $\Omega = d x^{1} 
\wedge \cdots \wedge d x^{n}.$

All integral elements on which $\Omega$ {\it does not vanish} are 
called {\bf admissible integral elements} according to \cite{Yang} and 
using this concept we can define
\bd{Involutive Systems}\\
An EDS with independence condition 
$(\Sigma,\Omega)$ is in involution with respect to $\Omega$ 
at a point $x \in \mathcal{M}$ means that there exists an {\bf admissible 
ordinary} integral element of dimension $n$ of $(\Sigma, \Omega)$ at x.
\ed
For practical reasons new integers called the {\bf reduced Cartan characters}
are introduced so as to decide whether a given EDS $\Sigma$ is 
in involution or not with respect to an independence condition $\Omega$. 
In order to determine them, we need to introduce the {\bf reduced 
polar systems $H^{red}((E^{p})_{x})$} which are defined as 
the polar systems $H^{red}((E^{p})_{x})$, where all terms involving any of the
$\omega^{1},\cdots ,\omega^{n}$ are suppressed. We define the reduced rank 
$s_{0}'(x):=r(H^{red}((E^{0})_{x}))$ and $s_{0}':=max(_{\{ x \in \mathcal{M} 
\} }s_{0}'(x))$ so that we can define the reduced characters inductively.
\bd{Reduced Cartan Characters}\\
The reduced Cartan characters $s_{p}^{\prime}$ are inductively defined as$$
s_{p}':=max(_{\{ x \in \mathcal{M}, V^{p+1}_{x} \in T_{x}\mathcal{M}\} }
      \: r(H^{red}((E^{p})_{x}))) -\sum_{i=0}^{p-1}s_{i}' \: .$$
\ed
The coincidence of the set of characters with that of the set of reduced 
characters is a necessary condition for a system to be in involution.
In practice, we can use {\bf Cartan's test for involution} based on a 
comparison  of polar elements with the reduced polar elements. But first, we 
must enlarge our account of Pfaffian systems with the use of more general 
concepts from the literature. Given a Pfaffian system $\mathcal{P}$ 
consisting of $s$ 1-forms $\theta^{\alpha}$ with independence condition 
$\Omega=\omega^{1} \wedge \cdots \wedge \omega^{n} \neq 0$, we denote by 
$\pi^{\lambda}$ all the extra forms such that 
$(\theta^{\alpha},\omega^{i}, \pi^{\lambda}), \: \: 1 \leq \alpha \leq s,$ 
form a coframe on our formal $N$-dimensional manifold $\mathcal{M}$. Once we
have chosen such forms $\pi^{\lambda}$, we can write each $d \theta^{\alpha}$ 
as
\be
d \theta^{\alpha} = A^{\alpha}_{\lambda i}\pi^{\lambda} \wedge 
\omega^{i} +\frac{1}{2}B^{\alpha}_{ij}\omega^{i}\wedge \omega^{j}
+\frac{1}{2}C^{\alpha}_{\lambda \kappa}\pi^{\lambda}\wedge 
\pi^{\kappa} \: (\mbox{mod} \: \mathcal{I}(\mathcal{P})) \label{2Pfacof} \: .
\ee
In equations (\ref{2Pfacof}), the $A^{\alpha}_{\lambda i}$ form the 
{\bf tableau matrix} and the $B^{\alpha}_{ij}$ are called the 
{\bf torsion terms}\footnotemark[1].
The Pfaffian system is called {\bf quasi-linear} if all the 
$C^{\alpha}_{\lambda \kappa}=0$. The following changes of coframe 
preserve linearity
\bear
\tilde{\theta}^{\alpha} & = & T^{\alpha}_{\beta}\theta^{\beta} \nn\\
\tilde{\omega}^{i} & = & T^{i}_{j}\omega^{j}+T^{i}_{\alpha}\theta^{\alpha}
\nn\\
\tilde{\pi}^{\lambda} & = & T^{\lambda}_{\kappa}\pi^{\kappa} 
+ T^{\lambda}_{j}\omega^{j}+T^{\lambda}_{\alpha}\theta^{\alpha} 
\label{2Pfaffch} \: .
\eear
We say that the torsion can be absorbed if a suitable transformation $\Phi$ 
$$\Phi : \pi^{\lambda} \rightarrow \pi^{\lambda} 
+ p^{\lambda}_{i}\omega^{i} $$
with
\be
{\tilde{B}}^{\alpha}_{ij} = B^{\alpha}_{ij}+ A^{\alpha}_{\lambda j}
p^{\lambda}_{i}-A^{\alpha}_{\lambda i}p^{\lambda}_{j} 
\label{2Torsi}
\ee
can be found such that ${\tilde{B}}^{\alpha}_{ij}=0$. 
When $B^{\alpha}_{ij}\neq 0$ in every coframe $(\theta^{\alpha},\omega^{e},
\pi^{\lambda})$, the system possesses integrability conditions which will 
prevent our given Pfaffian system $\mathcal{P}$ from being in involution. If a 
transformation $\Phi$ above exists such that all above 
${\tilde{B}}^{\alpha}_{ij}=0$, then there are no integrability conditions 
to stop the system being involutive. 

We can rewrite (\ref{2Pfacof}) in the dual language of Vessiot vector fields.
If we choose a frame dual to the coframe $(\omega^{i},\theta^{\alpha},
\pi^{\lambda})$, which we denote by $(E_{i},F_{\alpha},G_{\lambda})$ such that
\bear
\omega^{i}(E_{j}) & = & \delta^{i}_{\: j} \: , \quad (i,j=1, \cdots , n) \nn\\
\theta^{\alpha}(F_{\beta}) & = & \delta^{\alpha}_{\: \beta} \: , \quad
(\alpha,\beta = 1, \cdots , s) \nn\\
\pi^{\lambda}(G_{\kappa}) & = & \delta^{\lambda}_{\: \kappa} \: , \quad
(\lambda, \kappa =1, \cdots N-n-s)
\eear 
two Vessiot vector fields $X,Y \in \mathcal{D}$ can then be recast as
\bear
X & = & X^{i}_{(\omega)}E_{i}+ X^{\alpha}_{(\theta)}F_{\alpha}
+ X^{\lambda}_{(\pi)}G_{\lambda} \: , \nn\\
Y & = & Y^{j}_{(\omega)}E_{j}+ Y^{\beta}_{(\theta)}F_{\beta}
+ Y^{\kappa}_{(\pi)}G_{\kappa} \: .
\eear
When we calculate $d \theta^{\alpha}(X,Y)$, we obtain that
\bear
d \theta^{\alpha} (X,Y) & = & A^{\alpha}_{\lambda i}(X^{\lambda}_{(\pi)}
Y^{i}_{(\omega)}-X^{i}_{(\omega)}Y^{\lambda}_{(\pi)})+B^{\alpha}_{ij}
X^{i}_{(\omega)}Y^{j}_{(\omega)}+C^{\alpha}_{\lambda \kappa}X^{\lambda}_{(\pi)}
Y^{\kappa}_{(\pi)} \nn\\
& = & 0 \quad \mbox{mod} \: (\mathcal{I}(\mathcal{P})) \: , \label{duell}
\eear
where (\ref{duell}) is a system of equations for $X,Y \in \mathcal{D}$
based on the dual approach to (\ref{2Pfacof}).

From the tableau matrix $A^{\alpha}_{\lambda i}$, we can determine the 
reduced Cartan characters directly \cite{Yang}, \cite{5man}, where for the 
ranks $r(A^{\alpha}_{\lambda 1}):=s_{1}'$ and so on. There are many 
versions of Cartan's original test for involution \cite{Cart} some of 
which one can find in \cite{5man} or in \cite{Yang}, where a version 
for Pfaffian systems is discussed. We state it as:
\bt{Cartan's Test for Involution}\\
A Pfaffian system $(\mathcal{P},\Omega)$ is in involution means that 
$s_{0}=s_{0}',s_{1}=s_{1}', \cdots ,s_{p}=s_{p}'$
\underline{and} a coframe transformation can be found such that {\bf all}
its torsion terms $B^{\alpha}_{ij}$ vanish identically.
\et
The {\bf involutive genus} $g$ determines the maximal dimension a regular 
integral manifold can adopt and is given by $g=N-\sum^{n}_{i=0} s_{i}$. 
Note that the notion of an involutory subsystem of a vector field systems is 
not the same as the one used in discussing differential forms. In order to 
obtain the equivalent of this, we must add an
independence condition and define what it means for a vector field
system to be in involution with respect to $x^{1}, \cdots ,
x^{n}$. The precise definition can be found in \cite{Stor}. We can also use 
Groebner bases in developing criteria for involutivity. This is carried out 
in many places for instance in a paper by Mansfield \cite{Mans}.

In our example (\ref{2Pfacor}), we choose $w^{1}:=d x$ and $w^{2}:=d y$ so that
$\Omega = d x \wedge d y$ whereas the $u$ and $U$ constitute the dependent 
variables. Then, the reduced Cartan characters happen to coincide with their 
full counterparts so that $s=s_{0}=s_{0}'=12$ while all higher reduced 
characters vanish. Because the 12 $\theta^{\alpha}$ in (\ref{2Pfacor}) 
together with $\omega^{1} = d x$ and $\omega^{2}= d y$ form a complete coframe
of dimension $N=14$ already, we do not need to add any $\pi^{\lambda}$ and 
the tableau matrices 
vanish identically with all components $A^{\alpha}_{\lambda i}=0$. This proves 
that $s_{i}'=0, \: \forall i > 0$. The two torsion terms $B^{1}_{12}$ and
$B^{2}_{12}$ vanish identically. Next, we must compute the torsion terms 
which do not vanish identically and they are
$$B^{3}_{12} = B^{4}_{12} = B^{5}_{12} = B^{6}_{12} = rt+RT+1-s^{2}-S^{2} \: .
$$
But for our system transforming polar into Cartesian coordinates we have
$t=T=0$ and $s^{2}+S^{2}=1$ so that all the torsion terms vanish
identically and the system is in involution, and, the involutive 
genus $g$ is given by $g= 14 -12 = 2$.
\subsection{Existence Theorems and Prolongation}

An important question is whether and when integral manifolds
exist and whether, given some initial data, they are unique. 
The {\bf Cartan - K\"{a}hler theorems} \cite{5man,Yang}
give some answers to these questions for real-analytic systems. They reduce
to the {\bf Cauchy-Kovalevskaya theorem} when the number of equations equals
the number of unknowns which is given by $m$. A version for first-order 
systems is given in \cite{Grfo} which is sufficient because every system of 
PDEs can be transformed into a first-order system of PDEs.

The first Cartan-K\"{a}hler theorem specifies under which conditions a 
unique integral manifold of dimension $p$ can be constructed from a 
$(p-1)$-dimensional one.
For the second Cartan-K\"{a}hler existence theorem, we need our EDS
$\Sigma$ to be given in {\bf normal form}. Then, the second theorem simply 
states that under certain conditions a whole chain of regular integral 
manifolds of increasing dimensions exists (see Appendix A). 

If there is no general solution for a given EDS $(\Sigma, \Omega)$, 
then there could be identities or even integrability conditions so that we 
must {\bf prolong} the system by adding in these conditions and by enlarging 
the space of our formal jet coordinates to higher order. Good instructive 
examples of prolongation methods were given in \cite{EsWa1,EsWa2} and an 
example of an infinite dimensional prolongation of vector fields can be found 
in \cite{Fin1}. Next, we shalldiscuss Cartan's classical approach for an EDS 
briefly. But if we are dealing with a system of PDEs, then prolongation means 
just adding the partial derivatives of order $q+1$ as new jet-coordinates and
supplementing our original system of PDEs with all partial derivatives 
of the equations in the system \cite{Olv1}, \cite{Sei1}.

Cartan's classical procedure \cite{Cart}, \cite{5man} or \cite{Yang}
starts with an EDS with independence condition $(\Sigma, \Omega)$ given as 
\bear
\Sigma_{0} & = & (F_{1},...,F_{k_{0}}), \: k_{0} \: \: 0-forms \nn\\
\Sigma_{1} & = & (\alpha^{(1)}_{1},...,\alpha^{(1)}_{k_{1}}), \: k_{1} \: \: 
1-forms \nn\\
\Sigma_{2} & = & (\alpha^{(2)}_{1},...,\alpha^{(2)}_{k_{2}}), \: k_{2} \: \: 
2-forms \nn\\
 & & \vdots 
\eear
and so on. We now complete $(\alpha^{(1)}_{i_{1}},\omega^{i})$ to
a coframe $(\alpha^{(1)}_{i_{1}},\omega^{i},\pi^{\lambda})$ on $\mathcal{M}$.
We then choose an admissible integral element $(E^{n})_{x}$ from the set of all
admissible integral elements over $\mathcal{M}$, which we denote by 
$\mathcal{V}(\Sigma,\Omega)\subset \mathcal{V}_{n}(\Sigma)\: ,$ on which 
$$\alpha^{(1)}_{i_{1}}= t_{i_{1},i}\omega^{i} =0 \quad \mbox{and} \quad
\pi^{\lambda}=t^{\lambda}_{i}\omega^{i}$$ has to hold for some 
$(t_{i_{1},i};t^{\lambda}_{i})$ which are our new coordinates 
on the fiber $G_{p}(\mathcal{M})_{x} \cong G(N,p)\: ,$ where 
$T_{x}(\mathcal{M})\cong \mathbb{R}^{p}$. Based on this we can 
define the first prolongation of $(\Sigma,\Omega)$ as
\bd{First Prolongation}\\
The first prolongation $\Sigma^{(1)}$ of $(\Sigma, \Omega)$ is formed
by\\
i) the original EDS $(\Sigma, \Omega)$ restricted to $\mathcal{V}
(\Sigma,\Omega)$ and\\ 
ii) the closed system on $\mathcal{V}(\Sigma, \Omega)$ given by
$$\tilde{\pi}^{\lambda} = \pi^{\lambda}-t^{\lambda}_{i}
\omega^{i}\: ,$$ 
and
$$d \tilde{\pi}^{\lambda}\: .$$ 
\ed
We essentially added an extra number of contact conditions and 
their exterior derivatives to the above EDS $(\Sigma, \Omega)$.  
Higher-order prolongations are defined accordingly and for a detailed
account on prolongations see \cite{5man}. The key result for prolongations 
in the real analytic case \cite{Kura} is given by the Cartan-Kuranishi theorem
\bt{Cartan-Kuranishi Theorem}\\
An EDS $(\Sigma,\Omega)$ which is not in involution and not 
inconsistent becomes either involutive or inconsistent 
after a {\bf finite} number of prolongations.
\et
An upper limit for the number of prolongations for
involutivity or inconsistency is given by a number $\hat{q}$ which can
be determined inductively (c.f. \cite{Pom2}, \cite{Sei1} or \cite{Kura}). In 
the next sections, we shallapply some of the above theory to the 
Riemann-Lanczos problems in 4 and 5 dimensions.\\

\underline{Concluding Remarks}:
We should point out that there is a quantity similar to the Cartan 
characters also introduced by Cartan \cite{Cart} called the 
{\bf degr\'{e} d'arbitraire}. Seiler derived a relation 
between this degree of arbitrariness and the Cartan characters for 
nonlinear, overdetermined higher-order systems of PDEs \cite{Sei2}.
Einstein introduced the {\bf strength} of a system of equations in his 
last unified field theory of electromagnetism and gravity in order to 
measure how strongly such a system restricts the solution. A discussion 
between him and Cartan on this subject is presented in 
\cite{CElet}. Su\'{e} finally compared 
Cartan's degree of arbitrariness and Einstein's strength in 
\cite{Sue1,Sue2} for linear systems of equations.
\section{The Riemann-Lanczos Problem as an EDS and Singular Solutions}

In the Bampi-Caviglia papers \cite{Bam1,Bam2} the 
Riemann-Lanczos problem was written as an EDS and found not to be in 
involution in 4 dimensions as mentioned earlier. In their second paper 
\cite{Bam2} they introduced a prolongation to create an involutive system. 
Now, we are going to form the Pfaffian system resulting from the 
Riemann-Lanczos equations (\ref{1Rieml}). The Pfaffian system we are going to 
introduce slightly differs from that in \cite{Bam1} because we incorporate 
the cyclic conditions (\ref{1cyclic}) as well. The system consists of 20 
exterior derivatives of the Riemann-Lanczos equations together with the 20 
contact conditions and their exterior derivatives and we can omit the 
differential gauge conditions which do not change our results qualitatively. 
Using the exterior differentials of equations (\ref{Riemso}) in solved form
together with the contact conditions, we obtain the system
\bear
d f^{(R)}_{abcd} & = & (R_{abcd,e}+\alpha_{abcde})d x^{e}-d P_{abcd}
+d P_{abdc}-d P_{cdab}+d P_{cdba} \nn\\
& & +\G^{n}_{ad}(d L_{nbc}+d L_{ncb})-\G^{n}_{ac}(d L_{nbd}+d L_{ndb})\nn\\
& & +\G^{n}_{bc}(d L_{nad}+d L_{nda})-\G^{n}_{bd}(d L_{nac}+d L_{nca})\nn\\
K_{abc} & = & d L_{abc}-P_{abce} d x^{e} \nn\\
d K_{abc} & = & d x^{e} \wedge d P_{abce} \: , \label{5REDS}
\eear
where $\alpha$ is given by 
\bear
\alpha_{abcde} & = & \G^{n}_{ad,e}(L_{nbc}+L_{ncb})-\G^{n}_{ac,e}
(L_{nbd}+L_{ndb})+\G^{n}_{bc,e}(L_{nad}+L_{nda}) \nn\\
& & -\G^{n}_{bd,e}(L_{nac}+L_{nca}) \: . 
\eear
The $(x^{i},L_{abc},P_{abcd})$ are local jet coordinates on 
$\mathcal{J}^{1}(\mathbb{R}^{4},\mathbb{R}^{20})$ with $P_{abcd}=
\frac{\pr L_{abc}}{\pr x^{d}}$ when projected onto our spacetime manifold, 
and the $K_{abc}$ are the 20 contact conditions. 
A Vessiot vector field $V \in \mathcal{D}$ is locally given as
\be
V = V^{e}\frac{\pr}{\pr x^{e}}+ V^{e}P_{abce}\frac{\pr}{\pr L_{abc}}
+V_{abcd}\frac{\pr}{\pr P_{abcd}} \: , \label{Vesrie}
\ee
where 20 of the 80 components $V_{abcd}$ are defined through the requirement
that
\be
d f^{(R)}_{abcd}(V) = 0 \: .
\ee
This leads to the 20 independent equations
\bear
V_{ \{ abcd \} } & = & V^{e}[R_{abcd,e}+ \alpha_{abcde}
+\G^{n}_{ad}(P_{nbce}+P_{ncbe})-\G^{n}_{ac}(P_{nbde}+P_{ndbe}) \nn\\
& & +\G^{e}_{bc}(P_{nade}+P_{ndae})-\G^{n}_{bd}(P_{nace}+P_{ncae})] \: ,
\label{sammi}
\eear
where $\{abcd \}$ denotes the Riemann-type symmetrisation performed over the 
indices $abcd$. The 60 remaining components $V_{abcd}-V_{\{abcd \} }$ can be 
chosen arbitrarily. As shown in \cite{Bam1} a rank deficiency between the 
$s_{i}$ and the $s_{i}'$ occurs leading to 6 identities and the system 
(\ref{5REDS}) fails to be involutive. A more detailed explanation on how 
those identities can be created as linear combinations of rows of polar space 
matrices can be found in \cite{Ger1}. 

The reduced Cartan characters can be computed using the REDUCE computer code 
in \cite{Ger1} based on the EDS package by Hartley \cite{Har1} and slight 
modifications of it. For a number of 
spacetimes such as conformally flat, Kasner, G\"{o}del, some Debever types 
and plane wave spacetimes we verified that the reduced Cartan characters 
$(s_{0}',s_{1}',s_{2}',s_{3}',s_{4}')$ were 
(40,20,19,15,6) when the differential gauge condition (\ref{1diffg}) was not 
included and (40,20,19,15,0) when it was included. In both cases, the systems 
were clearly not in involution because some of the reduced characters differ 
from the Cartan characters - a fact reflected by results given by the 
computer code. 

Even though the Riemann-Lanczos problem in 4 dimensions is not in involution,
singular solutions do exist. Singular solutions occur when some 
of the integral elements, which are tangent to an integral manifold, are 
{\it singular} leading to singular integral manifolds, which means that the 
Cartan characters of the corresponding integral elements are not maximal.

In the remaining sections of this paper, we shallgive a few singular 
solutions for the Riemann-Lanczos problem in 4 dimensions for a number of 
spacetimes and briefly discuss the Riemann-Lanczos problem in 5 
dimensions.
\subsection{Debever-type Spacetimes: Example}

We expect singular solutions to occur for most of the Debever type spacetimes
for which a line element is given by
\be
d s^{2} = d t^{2}- f^{2}(t,y,z) d x^{2}- d y^{2} -d z^{2}
\ee
with $f(t,y,z)$ being one of the functions given in \cite{Deb3} or in 
\cite{Brun}. Specifically, for a Debever spacetime with $f=y^{2}$ of
Petrov type $D$ we are interested in singular solutions such that  
\be
\pounds _{\xi} L_{abc} =0
\ee
is possible using $\pounds_{\xi}R_{abcd}=0$ and the fact that 
$\nabla \pounds_{\xi} = \pounds_{\xi} \nabla$ for $\xi$ any Killing vector 
field (=KV). For convenience we shalldenote the local coordinates by
$x^{1}:=t,x^{2}:=x,x^{3}:=y,x^{4}:=z$ for this example. Using CLASSI 
\cite{Aman} we can determine the dimension of the isometry group which is 4 
in this case. There are 3 KVs corresponding to the ignorable coordinates 
$t,x$ and $z$ when the above local coordinate frame is used and 
the components of the fourth KV are $\xi^{4}=(z,0,0,t)$. Imposing 
$\pounds_{\xi}L_{abc}=0$ on the $L_{abc}$ leads to 
\bear
L_{tzx} & = & 2 L_{txz} \: ,\nn\\
L_{tzy} & = & 2 L_{tyz} \: ,\nn\\
L_{txt} & = & L_{xzz} \: ,\nn\\
L_{tyt} & = & L_{yzz} \: . \label{2Debbe}
\eear
The only non-vanishing independent Riemann-Lanczos equations then are
$f^{(R)}_{txty},f^{(R)}_{xyxy},f^{(R)}_{xzyz}$: 
\bear
R_{txty}+2\G^{x}_{xy}L_{txt} - P_{txty} & = & 0\nn\\
\frac{1}{2}R_{xyxy}+\G^{x}_{xy}L_{xyx}-P_{xyxy} & = & 0\nn\\
R_{xzyz}+P_{xzzy}-2\G^{x}_{xy}L_{xzz} & = & 0 \: \label{Debb} .
\eear
One solution which does also satisfy the 6 components of the differential 
gauge condition (\ref{1diffg}) is given by
\[ \baar{ll}
L_{txt} = C_{1}y^{4} \: , \quad & L_{xyx} = C_{2}y^{2}
+y^{3} \: ,\nn\\
 & \nn\\
L_{tzt} = C_{3} \: , \quad & L_{txz} = C_{4} \: ,\nn\\
 & \nn\\
L_{tyz} = C_{5}y^{-2} \: , \quad & L_{xyy} = C_{6} \: ,
\eaar \]
where $C_{1}, \cdots , C_{6}$ are constants.
If we choose $C_{3}=C_{4}=C_{6}=0$ then only 3 independent components will
remain. On a submanifold with $f^{(R)}_{abcd}=0$ of (\ref{5REDS}) for the 
above line element this solution corresponds to 
a singular integral manifold for which the (reduced) characters are 
$s_{0}=s_{0}'=3$ while all higher characters vanish. Such an integral 
manifold is parameterised by the above expressions for the $L_{abc}$ and 
their corresponding $P_{abcd}$. The Vessiot vector fields which span the 
tangent spaces of this singular solution manifold can locally be given by
\bear
V^{1} & = & \frac{\pr}{\pr t} \: , \nn\\
V^{2} & = & \frac{\pr}{\pr x} \: , \nn\\
V^{3} & = & \frac{\pr}{\pr y}+ 4C_{1}y^{3}\frac{\pr}{\pr L_{txt}}
- 2C_{2}y^{-3}\frac{\pr}{\pr L_{tyz}}- 4C_{2}y^{-3} \nn\\
& & \frac{\pr}{\pr L_{tzy}}+ (2C_{3}y+3y^{2})\frac{\pr}{\pr L_{xyx}}
+ 4C_{1}y^{3}\frac{\pr}{\pr L_{xzz}} \nn\\
& & +12 y^{2}\frac{\pr}{\pr P_{txty}} 
+6C_{2}y^{-4}\frac{\pr}{\pr P_{tyzy}} 
+12C_{2}y^{-4}\frac{\pr}{\pr P_{tzyy}} \nn\\ 
& & +2(C_{3}+3y)\frac{\pr}{\pr P_{xyxy}}
+12y^{2}\frac{\pr}{\pr P_{xzzy}} \: , \nn\\
V^{4} & = & \frac{\pr}{\pr z} \: ,
\eear
where we substituted
\bear
V^{3}_{txty} & = & 12 y^{2} \: , \nn\\
V^{3}_{tyzy} & = & 6C_{2}y^{-4} \: , \nn\\
V^{3}_{xyxy} & = & 2(C_{3}+3y) \: , 
\eear
and where $V^{3}_{txty}=V^{3}_{xzzy}$ and $V^{3}_{tzyy}
=2V^{3}_{tyzy}$ because of (\ref{2Debbe}).
In the next section, we are going to dicuss examples of spacetimes where
$\pounds_{\xi}L_{abc}=0$ cannot be implemented but where other singular 
solutions can occur.
\subsection{Singular Solution for G\"{o}del Spacetime}

G\"{o}del spacetime is a perfect fluid spacetime admitting closed time-like 
curves with line element
\be
d s^{2} = a^{2}(d t^{2}-d x^{2}-d z^{2}+\frac{1}{2}e^{2x}d y^{2}
            +2 e^{x}d t d y) \label{Kasner} \: .
\ee
Again, we are going to replace the local coordinates in such a way that 
$x^{1}:=t, x^{2}:=x, x^{3}:=y, x^{4}:=z$. It is well known \cite{HawkEls} that 
this spacetime admits a $G_{5}$ as an isometry group of which 3 Killing 
vectors $\xi^{1},\xi^{2},\xi^{3}$ are based on the ignorable coordinates
$t,y,z$. The other two are given by
$$\xi^{4} = \frac{\pr}{\pr x}- y \frac{\pr}{\pr y}, \quad
\xi^{5} = -2e^{-x}\frac{\pr}{\pr t}+y \frac{\pr}{\pr x}
+(e^{-2x}-\frac{1}{2}y^{2})\frac{\pr}{\pr y} \: .$$
But here, the Riemann-Lanczos problem does not admit singular solutions which
can inherit all the spacetime symmetries. This is the case
because not all elements of the isometry group of the spacetime are symmetries
of the Riemann-Lanczos problem. 

However, it is possible to find singular solutions with $\pounds_{\xi}
L_{abc}=0$ imposed for the 3 Killing vectors based on ignorable coordinates.
An Ansatz where some components are proportional to exponentials $e^{x},
e^{2x}$ while all other components vanish \cite{Ger1} leads to the solution 
\[\baar{ll}
L_{txy} = -\frac{a^{2}}{8}e^{x}\: , & L_{tyx}=\frac{a^{2}}{8}e^{x}\: , 
\nn\\
 & \nn\\
L_{txt} = -\frac{a^{2}}{8} \: , & L_{xyy}=\frac{3a^{2}}{16}e^{2x} \: .
\eaar \]
This solution does not satisfy all the differential gauge 
conditions (\ref{1diffg}) however \cite{Ger1}. This singular solution manifold 
is again parameterised by the above components $L_{abc}$ and their derivatives
$P_{abcd}$. Its tangent spaces are spanned by 4 Vessiot vector fields which 
can locally be given as
\bear
V^{1} & = & \frac{\pr}{\pr t} \: , \nn\\
V^{2} & = & \frac{\pr}{\pr x} 
+\frac{a^{2}}{8}e^{x}\frac{\pr}{\pr L_{txy}}
+\frac{a^{2}}{8}e^{x}\frac{\pr}{\pr L_{tyx}}
+\frac{3 a^{2}}{8}e^{2x}\frac{\pr}{\pr L_{xyy}} \nn\\
& & +\frac{1}{3}e^{x}(\frac{5a}{16}-\frac{1}{2}-e^{2x}\frac{3a}{16})
\frac{\pr}{\pr P_{txyx}}-\frac{1}{3}e^{x}(\frac{5a}{16}-\frac{1}{2}
-e^{2x}\frac{3a}{16})\frac{\pr}{\pr P_{tyxx}} \nn\\
& & +\frac{3}{4a}e^{2x}\frac{\pr}{\pr P_{xyyx}} \: , \nn\\
V^{3} & = & \frac{\pr}{\pr y} \: , \nn\\
V^{4} & = & \frac{\pr}{\pr z} \: ,
\eear
where it is $V^{2}_{tyxx}=V^{2}_{txyx}$. 
This solution is singular and all Cartan characters vanish even 
$s_{0}=s_{0}'=0$ because no constants remain in the above solution.
\subsection{Kasner Spacetime}

Kasner spacetime admits a $G_{3}$ based on the 3 ignorable 
coordinates $x,y,z$, where we again replaced $x^{1}:=t,x^{2}:=x,x^{3}:=y,
x^{4}:=z$. If we impose $\pounds_{\xi}L_{abc}=0$, the 6 non-vanishing 
components of the Riemann-Lanczos equations result in
\bear
\fbox{1} \quad 0 & = & R_{txtx} + 2P_{txxt} - 2\frac{p_{1}}{t}L_{txx} \nn\\
\fbox{2} \quad 0 & = & R_{tyty} + 2P_{tyyt} - 2\frac{p_{2}}{t}L_{tyy} \nn\\
\fbox{3} \quad 0 & = & R_{tztz} + 2P_{tzzt} - 2\frac{p_{3}}{t}L_{tzz} \nn\\
\fbox{4} \quad 0 & = & R_{xyxy} - 2{t}^{2p_{2}-1}p_{2}L_{txx} 
- 2{t}^{2p_{1}-1}p_{1}L_{tyy} \nn\\
\fbox{5} \quad 0 & = &  R_{xzxz} - 2{t}^{2p_{3}-1}p_{3}L_{txx} 
- 2{t}^{2p_{1}-1}p_{1}L_{tzz} \nn\\
\fbox{6} \quad 0 & = & R_{xzxz} - 2{t}^{2p_{3}-1}p_{3}L_{tyy} 
- 2{t}^{2p_{2}-1}p_{2}L_{tzz} \label{5Kasni} \: .
\eear
We can now easily see that solving the last 3 equations of (\ref{5Kasni})
algebraically leaves us with a solution for $L_{txx}, L_{tyy},
L_{tzz}$. But, inserting this solution into the first 3 equations of
(\ref{5Kasni}) and solving for the $P_{abcd}$ leads to
\[ \baar {ll}
L_{txx} = -\frac{1}{4}p_{1}{t}^{2p_{1}-1}, & 
P_{txxt} = (\frac{1}{2}- \frac{3}{4}p_{1})p_{1}{t}^{2p_{1}-2}, \\
 & \\
L_{tyy} = -\frac{1}{4}p_{2}{t}^{2p_{2}-1}, & 
P_{tyyt} = (\frac{1}{2}- \frac{3}{4}p_{2})p_{2}{t}^{2p_{2}-2}, \\
 & \\
L_{tzz} = -\frac{1}{4}p_{3}{t}^{2p_{3}-1}, & 
P_{tzzt} = (\frac{1}{2}- \frac{3}{4}p_{3})p_{3}{t}^{2p_{3}-2} .
\eaar \]
We see that when we project this down onto the spacetime manifold, we obtain 
that $P_{txxt} \neq L_{txx,t}$, $P_{tyyt} \neq L_{tyy,t}$ and $P_{tzzt} \neq
L_{tzz,t}$ therefore {\bf not} being a solution to (\ref{1Rieml}). 
Therefore, for Kasner spacetime no singular solution inheriting 
$\pounds _{\xi}L_{abc}=0$ exists. But there are other singular 
solutions for which some of the components $L_{abc}$ are linear in 
either $x,y$ or $z$ with no Lie symmetries along Killing 
directions. We decide to make the Ansatz
\[ \baar{ll}
L_{txt} = C_{1}{t}^{n_{1}}x \:, \qquad & 
L_{txx} = C_{4}{t}^{n_{4}}\: , \nn\\
L_{tyt} = C_{2}{t}^{n_{2}}y\: , \qquad & 
L_{tyy} = C_{5}{t}^{n_{5}}\: , \nn\\
L_{tzt} = C_{3}{t}^{n_{3}}z\: , \qquad & 
L_{tzz} = C_{6}{t}^{n_{6}}\: , \nn\\
L_{xyx} = C_{7}{t}^{n_{7}}y\: , \qquad & 
L_{xyy} = C_{10}{t}^{n_{10}}x\: , \nn\\
L_{xzx} = C_{8}{t}^{n_{8}}z\: , \qquad & 
L_{xzz} = C_{11}{t}^{n_{11}}x\: , \nn\\
L_{yzy} = C_{9}{t}^{n_{9}}z\: , \qquad & 
L_{yzz} = C_{12}{t}^{n_{12}}y\: , \label{5Kassin} 
\eaar \]
where $C_{1},...,C_{12}$ and $n_{1},...,n_{12}$ are arbitrary constants.
Inserted into the Riemann-Lanczos equations, we obtain
\bear
\fbox{1} \quad 0 & = & 
R_{txtx}-2P_{txtx}+2P_{txxt}-2\frac{p_{1}}{t}L_{txx} \nn\\
\fbox{2} \quad 0 & = & 
R_{tyty}-2P_{tyty}+2P_{tyyt}-2\frac{p_{2}}{t}L_{tyy} \nn\\
\fbox{3} \quad 0 & = & 
R_{tztz}-2P_{tztz}+2P_{tzzt}-2\frac{p_{3}}{t}L_{tzz} \nn\\
\fbox{4} \quad 0 & = & 
R_{xyxy}-2P_{xyxy}+2P_{xyyx}-2{t}^{2p_{2}-1}p_{2}L_{txx}
-2{t}^{2p_{1}-1}p_{1}L_{tyy} \nn\\
\fbox{5} \quad 0 & = & 
R_{xzxz}-2P_{xzxz}+2P_{xzzx}-2{t}^{2p_{3}-1}p_{3}L_{txx}
-2{t}^{2p_{1}-1}p_{1}L_{tzz} \nn\\
\fbox{6} \quad 0 & = & 
R_{yzyz}-2P_{yzyz}+2P_{yzzy}-2{t}^{2p_{3}-1}p_{3}L_{tyy}
-2{t}^{2p_{2}-1}p_{2}L_{tzz} \nn\\
\fbox{8} \quad 0 & = & 
R_{txxy}+L_{tyt}t^{2p_{1}-1}p_{1}-L_{xyx}{t}^{-1}(p_{1}+2p_{2})
+P_{xyxt}\nn\\
\fbox{9} \quad 0 & = & 
R_{tyxy}-L_{txt}{t}^{2p_{2}-1}p_{2}-L_{xyy}{t}^{-1}
(2p_{1}+p_{2})+P_{xyyt}\nn\\
\fbox{13} \quad 0 & = & 
R_{txxz}+L_{tzt}{t}^{2p_{1}-1}p_{1}-L_{xzx}{t}^{-1}
(p_{1}+2p_{3})+P_{xzxt}\nn\\
\fbox{15} \quad 0 & = & 
R_{tzxz}-L_{txt}{t}^{2p_{3}-1}p_{3}-L_{xzz}{t}^{-1}
(p_{3}+2p_{1})+P_{xzzt}\nn\\
\fbox{17} \quad 0 & = & 
R_{tyyz}+L_{tzt}{t}^{2p_{2}-1}p_{2}-L_{yzy}{t}^{-1}
(p_{2}+2p_{3})+P_{yzyt}\nn\\
\fbox{19} \quad 0 & = & 
R_{tzyz}-L_{tyt}{t}^{2p_{3}-1}p_{3}-L_{yzz}{t}^{-1}
(p_{3}+2p_{2})+P_{yzzt} \label{5Kassys} \: ,
\eear
where we labelled the equations as explained in Appendix B.
Using the above Ansatz with $n_{1}=2p_{1}-2,n_{2}=2p_{2}-2,
n_{3}=2p_{3}-2,n_{4}=n_{1}+1,n_{5}=n_{2}+1,n_{6}=n_{3}+1$ and
$n_{7}=n_{10}=-2p_{3},n_{8}=n_{11}=-2p_{2},n_{9}=n_{12}=-2p_{1}$, 
we obtain a singular solution for Kasner spacetime with the following 
constants for the above Ansatz
\[ \baar{ll}
C_{4} = \frac{C_{1}}{(p_{1}-1)}-\frac{p_{1}}{2}\: , \qquad &
C_{5} = \frac{C_{2}}{(p_{2}-1)}-\frac{p_{2}}{2}\: , \nn\\
 & \nn\\
C_{6} = \frac{C_{3}}{(p_{3}-1)}-\frac{p_{3}}{2}\: , \qquad &
C_{7} = \frac{p_{1}C_{2}}{2-p_{1}}\: , \nn\\
 & \nn\\
C_{8} = \frac{p_{1}C_{3}}{2-p_{1}}\: , \qquad &
C_{9} = \frac{p_{2}C_{3}}{2-p_{2}}\: , \nn\\
 & \nn\\
C_{10} = \frac{-p_{2}C_{1}}{2-p_{2}}\: , \qquad &
C_{11} = \frac{-p_{3}C_{1}}{2-p_{3}}\: , \nn\\
 & \nn\\
C_{12} = \frac{-p_{3}C_{2}}{2-p_{3}}\: . &
\eaar \]
A computer code, which determines the rather longish expressions 
for $C_{1},C_{2},$\\
$C_{3}$ in terms of $p_{1},p_{2},p_{3}$, can be 
found in Appendix B or in \cite{Ger1}. For this solution all characters and 
their reduced counterparts vanish identically so that $s_{0}=s_{0}'=0$ 
because all constants are completely determined by $p_{1},p_{2},p_{3}$.
\section{Comment on the Riemann-Lanczos Problem in 5 Dimensions}

On a 5-dimensional spacetime, we can again state the
Riemann-Lanczos problem. The structure of the Riemann-Lanczos equations 
remain the same as (\ref{1Rieml}) but, we obtain a larger number, namely 50 
independent equations due to the 50 independent components of the
Riemann tensor in 5 dimensions. Again, we omit the here 10 components of 
the differential gauge condition ${L_{ab}}^{s}_{\: ;s}=0$ in what follows.

We also obtain 50 Lanczos components $L_{abc}$ but imposing the 
cyclic conditions (\ref{1cyclic}) reduces the number to 40 independent 
components. Looking at the dimension of the jet bundle $J^{1}(\mathbb{R}^{5},
\mathbb{R}^{20})$, we find 5 spacetime dimensions, 40 independent 
Lanczos components $L_{abc}$ and 200 independent $P_{abcd}$ totalling 
$N=245$ formal dimensions for the manifold $\mathcal{M}$. We also find the 
total number of independent 1-forms is given by $s \leq 90$, where we have up 
to 50 independent $d f^{(R)}_{abcd}$ and 40 contact conditions $K_{abc}$. The 
number of independent components of a Vessiot vector field $V$ is given by 
$p=N-s$ which here amounts to $p \geq 155$. Such a Vessiot vector field can 
again be written as
\be
V = V^{e}\frac{\pr}{\pr x^{e}} + V^{e}P_{abce}\frac{\pr}{\pr L_{abc}}
  + V_{abcd}\frac{\pr}{\pr P_{abcd}}
\ee
with $V_{abcd}= V_{\{abcd\}}$, where $\{ abcd \}$ indicates Riemann type 
symmetries over the indices $abcd$. The EDS is again 
given by (\ref{5REDS}), the only difference being the number of equations 
involved and the range of the indices.

We wish to find the maximal dimension of a possible involution of Vessiot 
vector fields. The first Vessiot vector field $V^{1}$ has at least 155 
free (=parametric) components since $\mbox{dim}(\mathcal{D})=p \geq 155$. But 
all additional Vessiot vector fields $V^{2}, \cdots , V^{5}$ need to 
satisfy a further 40 conditions of the form
\be
d K_{abc}(V^{i},V^{j})=0 \: ,
\ee
where $K_{abc}=d L_{abc}-P_{abce}d x^{e}$ are the contact conditions in local 
coordinates. This leaves the next three Vessiot vector fields with $V^{2}$ 
having at least 115, $V^{3}$ at least 75 and $V^{4}$ at least 35 
parametric components. The last condition for $V^{5}$ can cause problems 
however, as there may not be enough free components to form a $V^{5}$ in
some cases. If we wish to find a suitable $V^{5}$, we must have 
$s_{0}+s_{1}+s_{2}+s_{3}+s_{4} \leq 240$. We can illustrate this as follows:
\[ \baar{ll}
s_{0} \leq 90 & \Rightarrow N- s_{0} \geq 155 \: , \nn\\
s_{0} + s_{1} \leq 130 & \Rightarrow N- s_{0} -s_{1} \geq 115 \: , \nn\\
s_{0} + s_{1} + s_{2} \leq 170 & \Rightarrow N- s_{0} -s_{1}-s_{2} \geq 
75 \: , \nn\\
s_{0} + s_{1} +s_{2} + s_{3} \leq 210 & \Rightarrow N- s_{0} +s_{1} + s_{2}
+ s_{3} \geq 35 \: , \nn\\
s_{0} + s_{1} + s_{2} + s_{3} + s_{4} \leq 250 & \Rightarrow N- s_{0} - s_{1}
-s_{2} - s_{3} - s_{4} \geq -5 \: ,
\eaar \]
where the second column gives the number of free components for $V^{1},
V^{2}$ and so on. Note that the fact that the sum of all free components
$155 + 115 + 75 + 35 = 380 \geq 245$ is permissible because 380 is the minimal
total number of free {\it but not necessarily independent components} for
all $V^{1}, \cdots , V^{5}$.

If we abandoned the cyclic conditions for a moment, the numbers would be:
$N=250, s \leq 100, p \geq 150$. Then $V^{1}$ would have at least 150 
components left, $V^{2}$ at least 100 components, $V^{3}$ at least 50 
components and $V^{4}$ would have no parametric components left in the 
worst case when all the $s_{i}$ are maximal. There would not be any parametric 
components left to form $V^{5}$ . Therefore, in the generic case no 
$5$-dimensional involution of Vessiot vector fields exists. This means that in
the generic case we cannot even form a singular $5$-dimensional chain of 
integral elements but only a $4$-dimensional at most. Therefore, neither the
first nor the second of the Cartan-K\"{a}hler theorems hold in this case.

Apart from that, we expect 20 identities of the same kind as the 
identities in \cite{Bam1} to occur. Using the computer code in Appendix C 
given in \cite{Ger1} adapted to the 5-dimensional case, we could find the 
following results for the reduced Cartan characters for a number of spacetimes
such as Minkowski space, the 5-dimensional conformally flat spacetime
and the Debever spacetimes. Their reduced characters 
$(s_{1}',s_{2}',s_{3}',s_{4}',s_{5}')$ when not including the differential 
gauge condition (\ref{1diffg}) were (40,39,35,26,10) and the second set 
including the differential gauge conditions (\ref{1diffg}) was 
(40,39,35,26,0). The computer code also confirmed that neither of the systems 
were in involution. We conclude that in order to be a viable problem in 5 
dimensions, the Riemann-Lanczos problem needs to be modified significantly.
\section*{Conclusion}

We gave a review of the theory of exterior differential systems and Pfaffian
systems in particular. We illustrated this theory with the 2-dimensional 
example of a coordinate transformation expressed as a complete Pfaffian system.

Singular solutions for the Riemann-Lanczos problem in 4 dimensions do exist 
and we found some for Kasner, G\"{o}del and for a Debever type spacetime.
 
For the Riemann-Lanczos problem in 5 dimensions, 20 internal identities
of the same type as those in 4 dimensions can occur as a computer code 
suggests and a 5-dimensional involution of Vessiot vector fields
cannot be achieved in a generic case.
\section*{Acknowledgements}
Both authors thank D Hartley for valuable discussions as well as Prof L S
Xanthis. 

A Gerber would like to thank the Swiss National Science Foundation 
(SNSF) and the Dr Robert Thyll-D\"{u}rr Foundation. 
\renewcommand{\theequation}{A.\arabic{equation}}
\setcounter{equation}{0}
\section*{Appendix A: The Cartan-K\"{a}hler Existence Theorems}

The Cartan-K\"{a}hler existence theorems \cite{5man,Yang,Kaeh,Cart} are
stated here. The first Cartan-K\"{a}hler theorem specifies under which 
conditions a unique integral manifold of dimension $(p+1)$ can be found from
a $p$-dimensional one:
\bt{First Cartan-K\"{a}hler Existence Theorem}\\
Let $\mathcal{M}$ be a formal $M$-dimensional manifold and $\mathcal{N}$ be a 
$p$-dimensional integral manifold with a regular integral element 
$T_{x}(\mathcal{N})$, where $(E^{p})_{x} = T_{x}(\mathcal{N})$ at a point $x$ 
on $M$. Further, there exists a submanifold $\mathcal{F}$, where
$\mathcal{N} \subset \mathcal{F} \subset \mathcal{M}$ such that 
$\mbox{dim}(\mathcal{F})=M-t_{p+1}$, $\mbox{dim}(T_{x}(\mathcal{F})\cap 
H((E^{p})_{x}))=p+1$, where $t_{p+1}= \mbox{dim}(H((E^{p})_{x}))-p-1$. 
Then, there exists a unique integral manifold $\mathcal{X}$ around $x$ 
such that $\mbox{dim}(\mathcal{X})=p+1$ and $\mathcal{F} \supset \mathcal{X} 
\supset \mathcal{N}$.
\et
This means that for a given $p+1$-dimensional manifold $\mathcal{F}$, we are
looking for the submanifold $\mathcal{X} \subset \mathcal{F}$ such that
$\mbox{dim} (\mathcal{X})= p+1$ and the theorem tells us that $\mathcal{X}$ 
exists and is unique.
Here, the integer $t_{p+1}$ is defined as $t_{p+1}=dim(E^{p})_{x}-p-1 \: ,
\: 0 \leq p < n$ and $s_{p}=t_{p}-t_{p+1}-1$. For $t_{0}$ we have 
$t_{0}=\sum_{i=0}^{n}s_{i}+n=dim \mathcal{I}(\Sigma_{0})\: .$
For the second Cartan-K\"{a}hler existence theorem, we need our EDS
$\Sigma$ to be given in {\bf normal form} which in this context means 
that there exists a local coordinate system $(x^{1},..,x^{n},y^{1},..,y^{r})$,
where $r = N-n$, such that:\\
\[ \baar{ll}
i) & \mathcal{I}(\Sigma_{0}) \: \mbox{is defined by} \: 
y^{t_{0}-n+1}= \cdots = y^{N-n} = 0 \: ,\\
ii) & H((E^{p})_{x})=\{\pr_{x^{1}}, \cdots ,\pr_{x^{n}},
\pr_{y^{(s_{0}+ \cdots + s_{p-1}+1)}}, \cdots ,\pr_{y^{(t_{0}-n)}})\}, 
\: 0 \leq p < n \: ,\\
iii) & (E^{p})_{x}=(\pr_{x^{1}}, \cdots ,\pr_{x^{p}}), 1 \leq p \leq n
\: , \\
iv) & \mbox{The integral point} \: (E^{0})_{x} \: \mbox{is given by} 
 (E^{0})_{x}=(0,\cdots,0) \: , \mbox{so that the point} \nn\\ 
& \mbox{$x$ coincides with} (0,\cdots,0) \mbox{($N$-times) on} \: 
\mathcal{M}\: .
\eaar \]
Then, we can state the second Cartan-K\"{a}hler existence theorem:
\bt{Second Cartan-K\"{a}hler Existence Theorem}\\
Given a regular chain of integral elements $(E^{0})_{x} \subset \cdots \subset
(E^{n})_{x}$ of an EDS $\Sigma$ given in normal form. 
Consider a set of initial data\\ 
$f_{1},\cdots,f_{s_{0}}\:$, ($s_{0}$ arbitrary constants), \\
$f_{s_{0}+1}(x^{1}),\cdots,f_{s_{0}+s_{1}}(x^{1})\:$, ($s_{1}$ 
arbitrary functions of 1 variable each), \\
$\vdots$\\
$f_{s_{0}+ \cdots +s_{n-1}+1}(x^{1},\cdots,x^{n}),\cdots,f_{t_{0}-n}
(x^{1},\cdots,x^{n})\:$, ($s_{n}$ arbitrary functions of n variables
$x^{1},\cdots,x^{n})\: .$ \\
For sufficiently small values of all the above $f_{i}$ and their
first-order derivatives there exists a unique integral manifold
defined by $y^{i}(x^{1},\cdots ,x^{n}),
y^{j}=0, 1\leq i \leq t_{0}-n \le j \leq N-n$, such that
$y^{i}(x^{1},\cdots,x^{p},0,\cdots,0)=f_{i}(x^{1},\cdots,x^{p})$
for $s_{0}+\cdots s_{p-1} < i \leq s_{0}+ \cdots + s_{p}, \:\: 0
\leq p \leq n.$   
\et
The highest non-vanishing Cartan character say $s_{k}, \: k \leq n$ is an 
{\bf invariant} of an EDS and its value gives the number of arbitrary 
functions of $k$ variables involved in the general solution. The other 
characters have the same meaning {\bf only if we used normal local
coordinates in the above sense}.
\renewcommand{\theequation}{B.\arabic{equation}}
\setcounter{equation}{0}
\section*{Appendix B: Labelling of the $f^{(R)}_{abcd}$ and Constants for 
Kasner Spacetime}

The labelling of the equations for the Riemann-Lanczos problem in 4 
dimensions as used in section III.C was carried out in the following way:
\[ \baar{llll}
\fbox{1} \leftrightarrow f^{(R)}_{1212} \: , &
\fbox{2} \leftrightarrow f^{(R)}_{1313} \: , &
\fbox{3} \leftrightarrow f^{(R)}_{1414} \: , &
\fbox{4} \leftrightarrow f^{(R)}_{2323} \: , \nn\\
 & & & \nn\\
\fbox{5} \leftrightarrow f^{(R)}_{2424} \: , & 
\fbox{6} \leftrightarrow f^{(R)}_{3434} \: , &
\fbox{7} \leftrightarrow f^{(R)}_{1213} \: , &
\fbox{8} \leftrightarrow f^{(R)}_{1223} \: , \nn\\
 & & & \nn\\
\fbox{9} \leftrightarrow f^{(R)}_{1323} \: , &
\fbox{10} \leftrightarrow f^{(R)}_{1324} \: , &
\fbox{11} \leftrightarrow f^{(R)}_{1214} \: , & 
\fbox{12} \leftrightarrow f^{(R)}_{1314} \: , \nn\\
 & & & \nn\\
\fbox{13} \leftrightarrow f^{(R)}_{1224} \: , &
\fbox{14} \leftrightarrow f^{(R)}_{2324} \: , &
\fbox{15} \leftrightarrow f^{(R)}_{1424} \: , &
\fbox{16} \leftrightarrow f^{(R)}_{1234} \: , \nn\\
 & & & \nn\\
\fbox{17} \leftrightarrow f^{(R)}_{1334} \: , & 
\fbox{18} \leftrightarrow f^{(R)}_{2334} \: , &
\fbox{19} \leftrightarrow f^{(R)}_{1434} \: , &
\fbox{20} \leftrightarrow f^{(R)}_{2434} \: .
\eaar \]
Next, we would like to give the computer code used to determine the 
constants for the singular solution for Kasner spacetime. The expressions for
the constants $C_{4}, \cdots , C_{12}$ given in section $III.C$ above can also
be found in \cite{Ger1}. They will be used as input for the REDUCE code which 
determines $C_{1},C_{2},C_{3}$ completely in terms of $p_{1},p_{2},p_{3}$. 
This code is given by:
\begin{verbatim}
%Constants for a singular solution for the Riemann-Lanczos
%problem for Kasner spacetime: 
%p1+p2+p3=1; p1**2+p2**2+p3**2=1;
%First we solve equations (1) to (3) and (7) to (12) and obtain:
C4:=C1/(p1-1)-p1/2;
C5:=C2/(p2-1)-p2/2;
C6:=C3/(p3-1)-p3/2;
C7:=(C2*p1/(2*p3+2*p2+p1));
C8:=(C3*p1/(2*p3+2*p2+p1));
C9:=(C3*p2/(2*p3+2*p1+p2));
C10:=(-C1*p2/(2*p1+2*p3+p2));
C11:=(-C1*p3/(p3+2*p2+2*p1));
C12:=(-C2*p3/(p3+2*p2+2*p1));
%We write (4), (5) and (6) for the remaining C1, C2 and C3 and
%solve:
s4:=2*C7-2*C10+2*p2*C4+2*p1*C5+p1*p2;
s5:=2*C8-2*C11+2*p3*C4+2*p1*C6+p1*p3;
s6:=2*C9-2*C12+2*p3*C5+2*p2*C6+p2*p3;
solve(s4=0,C1);
C1:=rhs first ws;
solve(s5=0,C2);
C2:=rhs first ws;
solve(s6=0,C3);
C3:=rhs first ws;

off nat,nero,echo;
out "H:/constants";
C1:=C1; C2:=C2; C3:=C3;
shut "H:/constants";
on nat,nero,echo;
end;
\end{verbatim}
The computer code above produces an output file called ``constants'' with the 
constants $C_{1}, C_{2}$ and $C_{3}$ expressed in terms of the $p_{i}$.
\vspace{0.3cm}
\footnoterule
\vspace{0.2cm}\noindent
$^{1}$ The term {\it torsion} in the context of EDS has nothing to do 
with the torsion occurring in the theory of affine connections.
\bibliography{refs}
\end{document}